\documentclass[fleqn,10pt]{article}
\usepackage{epsfig}
\newcommand{\be}{\begin{equation}}
\newcommand{\ee}{\end{equation}}

\newcommand{\eq}[1]{Eq.\,(\ref{#1})}
\newcommand{\fig}[1]{Fig.\,\ref{#1}}

\newcommand{\pbar}{\bar p}

\newcommand{\alphabold}{\mbox{\small\boldmath $\alpha$}}
\newcommand{\xbold}{\mbox{\boldmath $x$}}

\newcommand{\delchisq}{\Delta \chi^2_i(x_i;\alphabold)}
\newcommand{\delchi}{\Delta \chi^2_i}

\newcommand{\delchimax}{{\delchi}_{\rm max}}
\newcommand{\x}{\nu/m)}
\newcommand{\y}{(\nu_0/m)}
\begin{document}
\setcounter{secnumdepth}{4}
\renewcommand\thepage{\ }
%
%
\begin{titlepage} 
%
\newcommand\reportnumber{989} 
\newcommand\mydate{May 8, 2005} 
\newlength{\nulogo} 
\settowidth{\nulogo}{\small\sf{UW Report  MADPH-04-XXXX}}
\title{
\vspace{-.8in} 
\hfill\fbox{{\parbox{\nulogo}{\small\sf{
NUHEP Report  \reportnumber\\
UW Report MADPH-04-1382\\
          \mydate}}}}
\vspace{0.5in} \\
{
New evidence for the saturation of the Froissart bound 
}}

\author{
M.~M.~Block\\
{\small\em Department of Physics and Astronomy,} \vspace{-5pt} \\ 
{\small\em Northwestern University, Evanston, IL 60208}\\
\vspace{-5pt}
\  \\
F.~Halzen
\vspace{-5pt} \\ 
{\small\em Department of Physics,} 
\vspace{-5pt} \\ 
{\small\em University of
Wisconsin, Madison, WI 53706} \\
\vspace{-5pt}\\
%
\vspace{-5pt}\\
%
}    
\vspace{.5in}
\vfill
\date {}
\maketitle
\begin{abstract}
Fits to high energy data alone cannot cleanly discriminate between asymptotic $\ln s$ and $\ln^2s$ behavior of total hadronic cross sections. We demonstrate that this is no longer true when we require that these amplitudes {\em also} describe, on average, low energy data dominated by resonances. We simultaneously fit real analytic amplitudes to high energy measurements of: 1) the $\pi^+ p$ and $\pi^-p$ total cross sections and $\rho$-values (ratio of the real to the imaginary portion of the forward scattering amplitude), for $\sqrt s\ge 6$ GeV, while requiring that  the asymptotic fits smoothly  join the $\sigma_{\pi^+ p}$ and $\sigma_{\pi^-p}$ total cross sections at $\sqrt s=$2.6 GeV---{\em both} in magnitude and slope  , and 2) separately simultaneously fit the $\bar p p$ and $pp$ total cross sections and $\rho$-values  for $\sqrt s\ge 6$ GeV, while requiring that their asymptotic fits smoothly  join  the   the $\sigma_{\bar p p}$ and $\sigma_{pp}$ total  cross sections at $\sqrt s=$4.0 GeV---again {\em both} in magnitude and slope. In both cases, we have used all of the extensive data of the PDG group\cite {pdg}. However, we then subject these data to a screening process, the ``Sieve'' algorithm\cite{sieve}, in order to eliminate ``outliers'' that can skew a $\chi^2$ fit. With the ``Sieve'' algorithm, a robust fit using a Lorentzian distribution is first made to all of the data to sieve out abnormally high $\delchi$, the individual i$^{\rm th}$ point's contribution to the total $\chi^2$. The $\chi^2$ fits are then made to the sieved data. Both the  $\pi p$ and nucleon-nucleon systems strongly favor  a high energy $\ln^2s$ fit of the form: $\sigma^{\pm}=c_0 +c_1{\ln }\left({\nu\over m}\right)+c_2{\ln }^2\left({\nu\over m}\right)+\beta_{\cal P'}\left({\nu\over m}\right)^{\mu -1}\pm \delta\left({\nu\over m}\right)^{\alpha -1}$, basically excluding a $\ln s$ fit of the form: $\sigma^{\pm}=c_0 +c_1{\ln }\left({\nu\over m}\right)+\beta_{\cal P'}\left({\nu\over m}\right)^{\mu -1}\pm\delta\left({\nu\over m}\right)^{\alpha -1}$. The upper sign is for $\pi^+p$ ($pp$) and the lower sign is for $\pi^-p$ ($\bar pp$) scattering,   where $\nu$ is the laboratory  pion (proton) energy, and $m$ is the pion (proton) mass. 
\end{abstract}
\end{titlepage} 
%
\pagenumbering{arabic}
\renewcommand{\thepage}{-- \arabic{page}\ --}  
High energy cross sections for the scattering of hadrons should be bounded   by $\sigma \sim \ln^2s$,
where $s$ is the square of the cms energy.   This fundamental result is derived from unitarity and analyticity by Froissart\cite{froissart}, who states: ``At forward or backward angles, the modulus of the amplitude behaves at most like $s\ln^2s$, as $s$ goes to infinity.  We can use the optical theorem to derive that the total cross sections behave at most like $\ln^2s$, as $s$ goes to infinity".  In this context, saturating the Froissart bound refers to an energy dependence of the total  cross section rising no more rapidly than  $\ln^2s$.

The question as to whether any of the present day  high energy data for  $\bar pp$, $pp$ and $\pi^+ p$, $\pi^-p$  cross sections saturate the Froissart bound has not been settled; one can not unambiguously  discriminate between asymptotic fits of $\ln s$ and $\ln^2 s$  using high energy data only\cite{bkw,cudell}.  We here point out that this ambiguity is resolved by requiring that the fits to the high energy data smoothly join the cross section and energy dependence obtained by averaging the resonances at low energy. Imposing this duality\cite{igi} condition, we show that only fits to the high energy data behaving as $\ln^2 s$ that smoothly join (in {\em both} magnitude and first derivative) to the low energy data at the ``transition energy" (defined as the energy region just after the resonance regions end)  can  adequately describe the highest energy points.  This technique has recently been successfully used by Block and Halzen\cite{BH} to show that the Froissart bound is saturated for the  $\gamma p$ system.

We will use real analytic amplitudes to describe the data. Following Block and Cahn\cite{bc}, we write the  crossing-even real analytic amplitude for high energy  scattering  as\cite{BH},\cite{compton}
\begin{equation}
f_+=i\frac{\nu}{4\pi}\left\{A+\beta[\ln (s/s_0) -i\pi/2]^2+cs^{\mu-1}e^{i\pi(1-\mu)/2}-i\frac{4\pi}{\nu}f_+(0)\right\},
\end{equation}
and the crossing-odd amplitude as 
\be
f_-=-Ds^{\alpha -1}e^{i\pi(1-\alpha/2}.
\ee
where $A$, $\beta$, $c$, $s_0$ and $\mu$ are real constants. The variable $s$ is the square of the center of mass system (cms) energy and $\nu$ is the laboratory momentum. The additional real constant $f_+(0)$ is the subtraction constant at $\nu=0$ needed to be introduced in a singly-subtracted dispersion relation\cite{bc},\cite{gilman}.  Using the optical theorem, we obtain the total cross section
\be
\sigma^\pm= A+\beta\left[\ln^2 s/s_0-\frac{\pi^2}{4}\right]+c\,\sin(\pi\mu/2)s^{\mu-1}\pm D\cos(\pi\alpha/2)s^{\alpha -1}  \label{sigmatot}
\ee
with  $\rho$, the ratio of the real to the imaginary part of the forward scattering amplitude,  given by
\be
\rho^\pm={1\over\sigma_{\rm tot}}\left\{\beta\,\pi\ln s/s_0-c\,\cos(\pi\mu/2)s^{\mu-1}+\frac{4\pi}{\nu} f_+(0)\pm D\sin(\pi\alpha/2)s^{\alpha -1}\right\},\label{rhogeneral}
\ee 
where the upper sign is for $\pi^+p$ ($pp$) and the lower sign is for $\pi^-p$ ($\bar p p$) scattering, and the even amplitude applies to the spin-averaged $\gamma p$ scattering\cite{BH}.  

We now introduce the definitions $A = c_0 + \frac{\pi^2}{4}c_2 - \frac{c_1 ^ 2}{ 4c_2}$, $ s_0 = 2m ^ 2 e^{-c_1 / (2c_2)}$, $\beta=c_2$, $c = \frac{(2m^2)^{1 - \mu} } {\sin(\pi\mu/ 2)}\beta_{\cal P'}$ and $D=\frac{(2m^2)^{1-\alpha}}{\cos(\pi\alpha/2)}\delta$. In the high energy limit, where $s\rightarrow2m\nu$,  
\eq{sigmatot} and \eq{rhogeneral}, along with their cross section derivatives $\frac{d\sigma^{\pm}}{d(\x}$, can be written as
\begin{eqnarray}
\sigma^\pm&=&c_0+c_1\ln\left(\frac{\nu}{m}\right)+c_2\ln^2\left(\frac{\nu}{m}\right)+\beta_{\cal P'}\left(\frac{\nu}{m}\right)^{\mu -1}\pm\  \delta\left({\nu\over m}\right)^{\alpha -1},\label{sigmapm}\\
\rho^\pm&=&{1\over\sigma^\pm}\left\{\frac{\pi}{2}c_1+c_2\pi \ln\left(\frac{\nu}{m}\right)-\beta_{\cal P'}\cot({\pi\mu\over 2})\left(\frac{\nu}{m}\right)^{\mu -1}+\frac{4\pi}{\nu}f_+(0)
\pm \delta\tan({\pi\alpha\over 2})\left({\nu\over m}\right)^{\alpha -1} \right\}\label{rhopm},\\
\frac{d\sigma^{\pm}}{d(\x}&=&c_1\left\{\frac{1}{(\x)}\right\} +c_2\left\{ \frac{2\ln(\x)}{(\x)}\right\}+\beta_{\cal P'}\left\{(\mu-1)(\x)^{\mu-2}\right\}\label{derivpm}\\
&&\ \ \ \ \ \ \ \ \ \ \ \ \ \ \ \ \ \ \ \ \  \pm \ \delta\left\{(\alpha -1)(\x)^{\alpha - 2}\right\},
\end{eqnarray}
where the upper sign is for $\pi^+p$ ($pp)$ and the lower sign is for $\pi^-p$ ($\bar p p$) scattering.
The exponents $\mu$ and $\alpha$ are real. This transformation linearizes  \eq{sigmapm} in the real  coefficients $c_0,c_1,c_2$,$\beta_{\cal P'}$ and $\delta$, convenient for a $\chi^2$ fit to the experimental total cross sections and $\rho$-values.  Throughout we will use units of $\nu$ and $m$ in GeV and cross section in mb, where $m$ is the projectile mass. 

Let  $\sigma^+$ be the total cross section for $\pi^+p$ ($pp$) scattering and $\sigma^-$  the total cross section for $\pi^-p$ ($\bar p p$)   scattering. 
It is convenient to define, at the transition energy $\nu_0$,  
\begin{eqnarray}
\sigma_{\rm av}&=&\frac{\sigma^{+}\y+\sigma^-\y}{2}\nonumber\\
&=&c_0+c_1\ln\y+c_2\ln^2\y+\beta_{\cal P'}\y^{\mu-1},\\
\Delta\sigma&=&\frac{\sigma^{+}\y-\sigma^-\y}{2}\nonumber\\
&=&\delta\y^{\alpha -1},\\
m_{\rm av}&=&\frac{1}{2}\left(\frac{d\sigma^{+}}{d(\x}+\frac{d\sigma^{-}}{d(\x}\right)_{\nu =\nu_0}\nonumber\\
&=&c_1\left\{\frac{1}{\y}\right\}+c_2\left\{ \frac{2\ln\y}{\y)}\right\}+\beta_{\cal P'}\left\{(\mu-1)\y^{\mu-2}\right\},\\
\Delta m&=&\frac{1}{2}\left(\frac{d\sigma^{+}}{d(\x}-\frac{d\sigma^{-}}{d(\x}\right)_{\nu =\nu_0}\nonumber\\
&=&\delta\left\{(\alpha -1)\y^{\alpha - 2}\right\}.
\end{eqnarray}
Using the definitions of $\sigma_{\rm av}$, $\Delta\sigma$, $m_{\rm av}$ and $\Delta m$, we now write the four constraint equations
\begin{eqnarray}
\beta_{\cal P'}&=&\frac{\y^{2-\mu}}{\mu -1}\left[m_{\rm av}-c_1\left\{\frac{1}{\y}\right\} -c_2\left\{\frac{2\ln\y}{\y}
\right\}\right],\label{deriveven}\\
c_0&=& \sigma_{\rm av}-c_1\ln\y-c_2\ln^2\y-\beta_{\cal P'}\y^{\mu-1},\label{intercepteven}\\
\alpha&=&1+\frac{\Delta m}{\Delta \sigma}\y,\label{derivodd}\\
\delta&=&\Delta \sigma\y^{1-\alpha}\label{interceptodd},
\end{eqnarray}
that utilize the two slopes and the two intercepts at the transition energy $\nu_0$, where we join on to the asymptotic fit. We pick $\nu_0$ as the (very low) energy just after which resonance behavior finishes. We use $\mu=0.5$ throughout, which is appropriate for a Regge-descending trajectory.  In the above, $m=m_p$ is the proton mass for the $\bar pp$ and $pp$ systems, while $m=m_\pi$ is the pion mass for the $\pi^-p$ and $\pi^+p$ systems.

Our strategy is to use the rich amount of low energy data to constrain our  high energy fit. At the transition energy $\nu_0$,  the cross sections $\sigma^+(\nu_0/m)$ and $\sigma^-(\nu_0/m)$, along with the slopes $\left(\frac{d\sigma^{+}}{d(\x}\right)_{\nu=\nu_0}$ and $\left(\frac{d\sigma^{-}}{d(\x}\right)_{\nu=\nu_0}$, are used to constrain the asymptotic high energy fit so that it matches the low energy data at the transition energy $\nu_0$.   We pick $\nu_0$  much below the energy at which we start our high energy fit, but at an energy safely above the resonance regions. Very local fits are made to the region about the energy $\nu_0$ in order to evaluate the two cross sections and their two derivatives at $\nu_0$ that are needed in the above constraint equations. We next impose the 4 constraint equations, Equations (\ref{deriveven}), (\ref{intercepteven}), (\ref{derivodd}) and (\ref{interceptodd}), which we use in  our $\chi^2$ fit to Equations {\ref{sigmapm} and \ref{rhopm}. For safety, we  start the  data fitting at an energy $\nu_{\rm min}$ appreciably higher than the transition energy.  The transition energies, with appropriate cross sections and slopes, are summarized in Table \ref{table:transitionparameters}, along with the minimum energies used in the asymptotic fits.

 We stress that the odd amplitude parameters $\alpha$ and $\delta$ and hence the odd amplitude itself is {\em completely determined} by the experimental values $\Delta m$ and $\Delta \sigma$ at the transition energy $\nu_0$. Thus, at {\em all} energies, the {\em differences} of the cross sections $\sigma^- -\sigma^+$ (from the optical theorem, the differences in the imaginary portion of the scattering amplitude) and the {\em differences} of the real portion of the scattering amplitude are completely fixed {\em before} we make our fit.  Further, for a $\ln^2s$ $(\ln s$) fit, the even amplitude parameters $c_0$ and $\beta_{\cal P}'$ are determined by $c_1$ and $c_2$ ($c_1$ only) along with the experimental values of $\sigma_{\rm av}$ and $m_{\rm av}$ at the transition energy $\nu_0$. In particular, for a $\ln^2s$ ($\ln s$) fit, we only fit the 3 (2) parameters $c_1$, $c_2$, and $f(0)$ ($c_1$ and $f_+(0))$.  Since the subtraction constant $f_+(0)$ only enters into the $\rho$-value determinations, only the 2  parameters  $c_1$ and $c_2$  of the original 7 are required for a $\ln^2s$  fit to the cross sections $\sigma^{\pm}$, which gives  us exceedingly little freedom in this fit---it is indeed very tightly constrained, with not much latitude for adjustment.  The cross sections  $\sigma^{\pm}$ for the $\ln s$ fit are even more tightly constrained, with only one adjustable parameter, $c_1$.

We now outline the adaptive Sieve  algorithm\cite{sieve} that minimizes the effect that ``outliers''---points with abnormally high contributions to $\chi^2$---have on a fit when they contaminate a data sample that is otherwise Gaussianly distributed.  Our fitting procedure consists of several steps:
\begin{enumerate}
\item{Make a robust fit  of {\em all} of the data (presumed outliers and all)\ by minimizing $\Lambda^2_0$, the Lorentzian squared with respect to $\alphabold$, where 
\be
\Lambda^2_0(\alphabold;\xbold)\equiv\sum_{i=1}^N\ln\left\{1+0.179\delchisq\right\},\label{lambda0}
\ee 
 with  $\alphabold=\{\alpha_1,\ldots,\alpha_M\}$ being the $M$-dimensional parameter space of the fit. $\xbold=\{{x_1,\ldots,x_N}\}$ represents the abscissa of the $N$ experimental measurements $\mbox{\boldmath $y$}=\{y_1,\ldots,y_N\}$ that are  being fit and $\delchisq\equiv \left(\frac{y_i-y(x_i;\alphabold)}{\sigma_i}\right)^2$ is the individual $\chi^2$ contribution of the $i^{\rm th}$ point,  where $y(x_i;\alphabold)$ is the theoretical value at $x_i$ and $\sigma_i$ is the experimental error. It is shown in ref. \cite{sieve} that for Gaussianly distributed data, minimizing $\Lambda^2_0$ gives, on average,  the same total $\chi^2_{\rm min}\equiv\sum_{i=1}^N \delchisq$ from \eq{lambda0} as that found in a conventional $\chi^2$ fit,  as well as  rms widths (errors) for the parameters that are almost the same as those found in a $\chi^2$ fit}. 

A quantitative measure of whether point $i$ is an  outlier, {\em i.e.,} whether it is ``far away'' from the true signal,  is the magnitude of its $\delchisq= \left(\frac{y_i-y(x_i;\alphabold)}{\sigma_i}\right)^2$. The reason for minimizing the Lorentzian squared is that this procedure gives the outliers much less weight $w$ in the fit ($w\propto$ 1/$\sqrt{\delchisq}$, for large $\delchisq$) than does a $\chi^2$ fit ($w\propto \sqrt{\delchisq}$), thus making the fitted parameters insensitive to  outliers and hence robust. For details, see ref. \cite{sieve}.

If $\chi^2_{\rm min}$ is satisfactory, make a conventional $\chi^2$ fit to get the errors and you are finished.   If $\chi^2_{\rm min}$ is not satisfactory, proceed to step 
 \ref{nextstep}.
\item {Using the above robust $\Lambda^2_0$ fit as the initial estimator for the theoretical curve, evaluate $\delchisq$, for each of the $N$ experimental points.}\label{nextstep}
\item A largest cut, $\delchisq_{\rm max}$, must now be selected. We start the process with $\delchisq_{\rm max}=9$. If any of the points have $\Delta \chi^2_i(x_i;\alphabold)>\delchisq_{\rm max}$, reject them---they fell through the ``Sieve''. The choice of $\delchisq_{\rm max}$ is an attempt to pick  the largest ``Sieve'' size (largest $\delchisq_{\rm max}$) that rejects all of the outliers, while minimizing the number of signal points  rejected. \label{redo}
\item Next, make a conventional $\chi^2$ fit to the sifted set---these data points are the ones that have been retained in the ``Sieve''. This  fit is used to estimate   $\chi^2_{\rm min}$.    Since the data set has been truncated by eliminating the points with $\delchisq>\delchisq_{\rm max}$, we must slightly renormalize the $\chi^2_{\rm min}$ found to take this into account, by the factor $\cal R$.  For $\delchimax=9,6,$ and 4, the factor $\cal R$ is given by 1.027, 1.140 and 1.291, whereas the fraction of the points that should survive this $\chi^2$ cut---for a Gaussian distribution---is 0.9973, 0.9857 and 0.9545, respectively. A plot of ${\cal R}^{-1}$ as a function of $\delchimax$ is given in Figure \ref{renorm}, which is taken from ref. \cite{sieve}.

If the renormalized $\chi^2_{\rm min}$, {\em i.e.,} ${\cal R}\times \chi^2_{\rm min}$ is acceptable---in the {\em conventional} sense, using the ordinary $\chi^2$ distribution probability function---we consider the fit of the data to the  model to be satisfactory  and proceed to the next step. If the renormalized $\chi^2_{\rm min}$ is not acceptable and $\delchisq_{\rm max}$ is not too small, we pick a smaller $\delchisq_{\rm max}$ and go back to step \ref{redo}. The smallest value of $\delchisq_{\rm max}$ that we used is $\delchisq_{\rm max}=4$.  

\item
From the  $\chi^2$ fit that was made to the ``sifted'' data in the preceding step, evaluate  the parameters $\alphabold$.
Next, evaluate the $M\times M$ covariance (squared error) matrix of the parameter space which was found in the $\chi^2$ fit. We find the new squared error matrix for the $\Lambda^2$  fit by multiplying the covariance matrix by the square of the factor $r_{\chi^2}$. From Figure \ref{renorm}, we find that  $r_{\chi^2}\sim 1.02,1.05$ and  1.11  for $\delchisq_{\rm max}=9$, 6 and 4,  respectively . The values of $r_{\chi^2}>1$ reflect the fact that a $\chi^2$ fit to the {\em truncated} Gaussian distribution that we obtain---after first making  a robust fit---has a rms (root mean square) width which is somewhat greater than the  rms width of the $\chi^2$ fit to the same untruncated distribution\cite{sieve}. 
\end{enumerate}

The application of a $\chi^2$ fit to the {\em sifted set} gives stable estimates of the model parameters $\alphabold$, as well as a goodness-of-fit of the data to the model when $\chi^2_{\rm min}$ is renormalized for the effect of truncation due to the cut $\delchisq_{\rm max}.$  One can now use conventional probabilities for $\chi^2$ fits, {\em i.e.,} the probability that $\chi^2$ is greater than ${\cal R}\times\chi^2_{\rm min}$, for the number of degrees of freedom $\nu$. Model parameter errors are found by multiplying the covariance (squared error) matrix of the conventional $\chi^2$ fit by the appropriate factor $(r_{\chi^2})^2$ for the cut $\delchisq_{\rm max}$.

Table \ref{table:pipfitnew} summarizes the results of our simultaneous fits to all of the available data from the Particle Data Group\cite{pdg} for  $\sigma_{\pi^+p}$, $\sigma_{\pi^-p}$, $\rho_{\pi^+p}$ and $\rho_{\pi^-p}$, using the 4 constraint equations with a transition energy $\sqrt s=2.6$ GeV and a minimum fitting energy of 6 GeV, after applying  the ``Sieve'' algorithm\cite{sieve}.  Three $\delchimax$ cuts, 4, 6 and 9, were made for $\ln^2(\nu/m_\pi)$ fits. There was considerable improvement in the renormalized $\chi^2$/d.f. going from $\delchimax=9$ to $\delchimax=6$.  However, there was  no improvement of the renormalized $\chi^2$/d.f. going from $\delchimax=6$ to $\delchimax=4$---indeed, it increased    from 1.294 to 1.364.  Since the errors also become substantially larger for the $\delchimax=4$ cut, we chose to use the values of the $\ln^2(\nu/m_\pi)$ fit with a $\delchimax=6$ cut.  This cut was therefore also used for the $\ln(\nu/m_\pi)$ fit. The probability of the fit for the data set using the $\delchimax=6$ cut was $\sim 0.02$, a somewhat low probability, albeit one that is often deemed acceptable in a fit with this  many degrees of freedom (d.f.=127). In contrast, the probability of the $\ln(\nu/m_\pi)$ fit using  the $\delchimax=6$ data set is $<<10^{-16}$ and is clearly ruled out, as is graphically demonstrated in \fig{fig:sigpip}. 

It should be noted that when using a $\ln^2(\nu/m_\pi)$ fit {\em before} imposing the ``Sieve'' algorithm,  a value of $\chi^2$/d.f.=3.472 for 152 degrees of freedom was found, compared to $\chi^2$/d.f.=1.294 for 127 degrees of freedom when using the $\delchimax=6$ cut. In essence, the ``Sieve'' algorithm eliminated 25 points with energies $\sqrt s\ge6$ GeV (2 $\sigma_{\pi^+p}$, 19 $\sigma_{\pi^-p}$, 4 $\rho_{\pi^+p}$), while changing the total renormalized $\chi^2$ from 527.8 to 164.3. These 25 points that were screened out had a $\chi^2$ contribution of 363.5, an average value of 14.5.  If the distribution had been Gaussian with no outliers, one would have expected about 2 points having $\delchi>6$, giving a total $\chi^2$ contribution slightly larger than 12, compared to the observed value of 363.5.  Thus, we see the effect of the ``Sieve'' algorithm in cleaning up the data sample by eliminating the outliers.

Next, we analyze the $\bar pp$ and $pp$ systems. Table \ref{table:ppfitnew} summarizes the results of our simultaneous fits to the available accelerator data  from the Particle Data Group\cite{pdg} for  $\sigma_{pp}$, $\sigma_{\bar pp}$, $\rho_{pp}$ and $\rho_{\bar pp}$, using the 4 constraint equations with a transition energy $\sqrt s=4$ GeV and a minimum fitting energy of 6 GeV, again using the ``Sieve'' algorithm. Two $\delchimax$ cuts,  6 and 9, were made for $\ln^2(\nu/m_p)$ fits. The probability of the fit for the cut $\delchimax=6$ was $\sim 0.2$, a very satisfactory probability for this  many degrees of freedom, and we chose this data set rather than the data set corresponding to the $\delchimax=9$ cut. As seen in Table \ref{table:ppfitnew}, the fitted parameters are very insensitive to this choice.
 The same data set ($\delchimax=6$ cut) was also used for the $\ln(\nu/m_p)$ fit. The probability of the $\ln(\nu/m_p)$ fit is $<<10^{-16}$ and is clearly ruled out. This is illustrated graphically in \fig{fig:sigmapp} and \fig{fig:rhopp}.

We note  that when using a $\ln^2(\nu/m_p)$ fit {\em before} imposing the ``Sieve'' algorithm,  a value of $\chi^2$/d.f.=5.657 for 209 degrees of freedom was found, compared to $\chi^2$/d.f.=1.095 for 184 degrees of freedom when using the $\delchimax=6$ cut. The ``Sieve'' algorithm eliminated 25 points with energies $\sqrt s\ge6$ GeV (5 $\sigma_{pp}$, 5 $\sigma_{\pbar p}$, 15 $\rho_{pp}$), while changing the total renormalized $\chi^2$ from 1182.3 to 201.4. These 25 points that were screened out had a $\chi^2$ contribution of 980.9, an average value of 39.2. For a Gaussian distribution, about 3 points with $\delchi>6$ are expected, with a total $\chi^2$ contribution of slightly more than 18 and {\em not} 980.9. Again, we see the effect of the ``Sieve'' algorithm in ridding the data sample of outliers.

Fig. \ref{fig:sigpip} shows the individual fitted cross sections (in mb) for $\pi^+p$ and $\pi^-p$ for  $\ln^2(\nu/m_\pi)$ and $\ln(\nu/m_\pi)$,  for the cut $\delchimax=6$, from Table \ref{table:pipfitnew} plotted against the cms energy, $\sqrt s$, in GeV. The data shown are the sieved data with $\sqrt s \ge 6$ GeV. The $\ln^2(\nu/m_\pi)$ fits with $\delchimax=6$ corresponding to the solid curve for $\pi^-p$ and the dash-dotted curve for $\pi^+p$,  are in excellent agreement with the cross section data. On the other hand, the $\ln(\nu/m_\pi)$ fits---the long dashed curve for $\pi^-p$ and the short dashed curve for $\pi^+p$---although they fit in the low energy region almost identically to the $\ln^2(\nu/m_\pi)$ fits---are very bad fits which clearly underestimate {\em all} of the high energy cross sections, leading to huge $\chi^2_{\rm min}$, and hence are ruled out. 

Figure \ref{fig:rhopip} shows the individual fitted $\rho$-values for $\pi^+p$ and $\pi^-p$ for  $\ln^2(\nu/m_\pi)$ and $\ln(\nu/m_\pi)$,  for the cut $\delchimax=6$, from Table \ref{table:pipfitnew}, {\em vs.} $\sqrt s$, the cms energy  in GeV. The data shown are the sieved data with $\sqrt s \ge 6$ GeV. The $\ln^2(\nu/m_\pi)$ fits with $\delchimax=6$, corresponding to the solid curve for $\pi^-p$ and the dash-dotted curve for $\pi^+p$,  reproduce  the data reasonably well. On the other hand, the $\ln(\nu/m_\pi)$ fits are  rather poor.  Again the $\rho$ data offer firm support for the Froissart bound fits, while ruling out $\ln(\nu/m_\pi)$ fits for the $\pi p$ system.

Figure \ref{fig:sigpipallenergies} shows an expanded scale of energies, in which {\em all} available $\pi p$ cross sections are shown, from threshold to the highest available energies. The dashed curve is the {\em even} amplitude pion cross section $\left (\frac{\sigma_{\pi^+p}+\sigma_{\pi^-p}}{2}\right )$ computed from our $\delchimax=6$ cut, whereas the solid curve is the result of a similar analysis for the spin-averaged (even) cross section\cite{BH} for $\gamma p$ scattering, rescaled by multiplying it by 210, a  familiar number from the vector dominance model.  It is most striking that these two independent curves are virtually indistinguishable in the entire energy interval in which experimental data are available, {\em i.e.,} $2\le\sqrt s\le 300$ GeV---a result most strongly supporting the vector dominance model.

 All known $\pi p$ cross section data  are plotted in Figure \ref{fig:blockhalzen&Igi}, which  compares our  analysis using the 4 constraint equations ($\delchimax=6$ from Table \ref{table:pipfitnew}) with the analysis of Igi and Ishida\cite{igi} which used finite energy sum rules (FESR) for their low energy data. They only fitted the even cross section, so we have plotted in Fig. \ref{fig:blockhalzen&Igi}(a) the even portion of our $\ln^2(\nu/m_\pi)$ fit as the solid curve.  It is seen to go smoothly through the {\em average} cross section, $\left (\frac{\sigma_{\pi^+p}+\sigma_{\pi^-p}}{2}\right )$, for pion-proton scattering.  The dashed-dot curve, using the FESR, is from Igi and Ishida\cite{igi}.  It does not go very smoothly through the average of the points, but rather goes much closer to  $\sigma_{\pi^+p}$ in the energy region from 10 to 30 GeV. Perhaps this is the result of their trying to fit {\em only} the even cross section, whereas we separately fit $\sigma_{\pi^+p}$ and $\sigma_{\pi^-p}$. We have plotted in Fig. \ref{fig:blockhalzen&Igi}(b) the even portion of our $\ln(\nu/m_\pi)$ fit as the dashed curve, with the FESR result being the dashed-dot-dot curve.  Clearly both curves rule out a $\ln(\nu/m_\pi)$ behavior. Both analyses strongly support a $\ln^2(\nu/m_\pi)$ behavior and thus a saturation of the Froissart bound for the $\pi p$  system. 

Figure \ref{fig:sigmapp} shows the individual fitted cross sections (in mb) for $ pp$ and $\bar pp$ for $\ln^2(\nu/m_p)$ and $\ln(\nu/m_p)$ for the cut $\delchimax=6$ in Table \ref{table:ppfitnew}, plotted against the cms energy, $\sqrt s$, in GeV. The data shown are the sieved data with $\sqrt s \ge 6$ GeV. The $\ln^2(\nu/m_p)$ fits to the data sample with $\delchimax=6$, corresponding to the solid curve for $\bar pp$ and the dash-dotted curve for $pp$,  are excellent, yielding a total renormalized $\chi^2=201.5$, for 184 degrees of freedom, corresponding to a fit probability of $\sim0.2$. On the other hand, the $\ln(\nu/m_p)$ fits to the same data sample---the long dashed curve for $\bar p p$ and the short dashed curve for $pp$---are very bad fits, yielding a total $\chi^2=2613.7$ for 185 degrees of freedom, corresponding to a fit probability of $<<10^{-16}$. In essence, the $\ln(\nu/m_p)$ fit clearly undershoots {\em all} of the high energy cross sections. The ability of nucleon-nucleon scattering to distinguish cleanly  between an  energy dependence of $\ln^2(\nu/m_p)$ and  an energy dependence of $\ln(\nu/m_p)$ is even more dramatic than the pion result.

Figure \ref{fig:rhopp} shows the individual fitted $\rho$-values for $pp$ and $\bar pp$ $\ln^2(\nu/m_p)$ and $\ln(\nu/m_p)$ from Table \ref{table:ppfitnew}, using $\delchimax=6$---plotted against the cms energy, $\sqrt s$,  in GeV. The data shown are the sieved data with $\sqrt s \ge 6$ GeV. The $\ln^2(\nu/m_p)$ fits,
 corresponding to the solid curve for $\bar pp$ and the dash-dotted curve for $pp$,  fit the data reasonably well. On the other hand, the $\ln(\nu/m_p)$ fits, the long dashed curve for $\bar pp$ and the short dashed curve for $pp$, are very poor fits, missing completely the precise $\rho_{\bar pp}$ at 546 GeV, as well as  $\rho_{\bar pp}$ at 1800 GeV. These results again strongly support  the $\ln^2(\nu/m_p)$ fits that saturate the Froissart bound and once again rule out $\ln (\nu/m_p)$ fits for the $\bar pp$ and $p p$ system.

A few remarks on our $\ln^2(\nu/m_p)$ asymptotic energy analysis for $pp$ and $\bar pp$ are in order. It should be stressed that we used {\em both} the CDF and E710/E811 high energy experimental cross sections at $\sqrt s=1800$ GeV in the $\ln^2(\nu/m_p)$ analysis, summarized in Table \ref{table:ppfitnew}, $\delchimax=6$ and shown in Figures \ref{fig:sigmapp} and \ref{fig:rhopp}.  Inspection of Fig. \ref{fig:sigmapp} shows that at $\sqrt s=1800$ GeV, our fit  effectively passes below the  cross section point of $\sim$ 80 mb (CDF collaboration).  In particular, to test the sensitivity of our fit to the differences between the highest energy accelerator $\bar pp$ cross sections from the Tevatron, we next {\em omitted completely} the CDF ($\sim$ 80 mb) point and refitted the data without it.  This fit, also using $\delchimax=6$,  had a renormalized $\chi^2$/d.f.=1.055, compared to 1.095 with the CDF point included.  Since you only expect, on average, a $\Delta\chi^2$ of $\sim 1$ for the removal of one point, the removal of the CDF point slightly improved  the goodness-of-fit. Moreover, the new parameters of the fit were only {\em very minimally} changed. As an example, the predicted value from  the new fit for the cross section at $\sqrt s=1800$ GeV---{\em without} the CDF point---was $\sigma_{\bar pp}=75.1\pm0.6$ mb, where the error is the statistical error due to the errors in the  fitted parameters.   Conversely, the predicted value from 
Table \ref{table:predictions}---which used {\em both} the CDF and the E710/E811 point---was $\sigma_{\bar pp}=75.2\pm0.6$ mb, virtually identical. Further, at $\sqrt s=14$ TeV (LHC energy),  the fit {\em without} the  CDF point had $\sigma_{\bar pp}=107.2\pm1.2$, whereas {\em including} the CDF point (Table \ref{table:predictions}) gave $\sigma_{\bar pp}=107.3\pm1.2$. 
Thus, within errors, there was practically no effect of either including or excluding the CDF point. The fit was determined almost exclusively by  the E710/E811 cross section---presumably because the asymptotic fit was locked into the low energy transition energy $\nu_0$, thus sampling the rich amount of lower energy data.    

Our result concerning the (un)importance of the CDF point relative to E710/E811 result is to be contrasted with the statement from the COMPETE Collaboration\cite{cudell} which emphasized that there is: ``the systematic uncertainty coming from the discrepancy between different FNAL measurements of $\sigma_{\rm tot}$",  which contribute large differences to their fit predictions at high energy, depending on which data set they use.  In marked contrast to our results, they conclude that their fitting techniques favor the CDF point. Our results  indicate that {\em both} the cross section and $\rho$-value of the E710/E811 groups are slightly favored. More importantly, we find virtually {\em no sensitivity} to high energy predictions when we do not use the CDF point and only use the E710/E811 measurements. Our method of fitting the data---by anchoring the asymptotic fit at the low transition energy $\nu_0$---shows that our high energy predictions are quasi-independent of the FNAL ``discrepancy'', leading us to believe that our high energy cross section predictions at both the LHC and at cosmic ray energies  are both robust and accurate.  In Table \ref{table:predictions}, we give  predictions---from our  $\ln^2(\nu/m_p)$ fit---for some values of $\sigma_{\bar pp}$ and $\rho_{\bar pp}$ at high energies. The errors quoted are due to the statistical errors of the fitted parameters $c_1$,  $c_2$ and $f_+(0)$ given in the $\delchimax=6$, $\ln^2(\nu/m_p)$ fit  of Table \ref{table:ppfitnew}.

In Fig. \ref{fig:sigppallenergies}, we show an extended energy scale, from threshold up to cosmic ray energies ($1.876\le \sqrt s\le 10^5 $ GeV), 
plotting all available $\bar pp$ and $pp$ cross sections, including cosmic ray  $pp$ cross sections inferred from  cosmic ray p-air experiments by Block, Halzen and Stanov\cite{BHS}. The solid curve is our result from Table \ref{table:ppfitnew} of the {\em even} cross section from  $\ln^2(\nu/m_p)$, $\delchimax=6$. The dashed-dot-dot curve is from an independent QCD-inspired eikonal analysis\cite{BHS} of the nucleon-nucleon system.  The agreement is quite remarkable---the two independent curves are virtually indistinguishable over almost 5 decades of cms energy, from $\sim 3$ GeV to 100 TeV.  Figure \ref{fig:sigppallenergies} clearly indicates that the $p p$ and $\bar pp$ cross section data  greater than $\sim3$ GeV can  be explained by a fit of the form 
$\sigma^\pm=c_0+c_1\ln\left(\frac{\nu}{m_p}\right)+c_2\ln^2\left(\frac{\nu}{m_p}\right)+\beta_{\cal P'}\left(\frac{\nu}{m_p}\right)^{\mu -1}\pm\  \delta\left({\nu\over m_p}\right)^{\alpha -1}$
 over an enormous energy range, {\em i.e.,} by a $\ln^2s$ saturation of the Froissart bound.

In Table \ref{table:predictions}, we make predictions of  total cross sections and $\rho$-values for $\bar pp$ and $pp$ scattering---in  the low energy regions covered by RHIC, together with the energies of the Tevatron and LHC as well as the high energy regions appropriate to cosmic ray air shower experiments.

We  give strong support to  vector meson dominance by showing that the {\em even} cross section from our fits for $\pi^+p$ and $\pi^-p$ data agrees exceedingly well  with a rescaled (multiplied by a factor of 210) $\sigma_{\gamma p}$ analysis done earlier by Block and Halzen\cite{BH}, when both cross sections have a $\ln^2s$ asymptotic behavior. 

In conclusion, we have demonstrated that the duality requirement that high energy cross sections smoothly interpolate into the resonance region strongly favors a $\ln^2s$ behavior of the asymptotic cross sections for both the $\pi p$ and nucleon-nucleon systems, in agreement with our earlier result for $\gamma p$ scattering\cite{BH}.  We conclude that the three hadronic systems, $\gamma p$, $\pi p$ and  nucleon-nucleon, {\em all} have an asymptotic $\ln^2s$ behavior, thus saturating the Froissart bound.

At 14 TeV, we predict  $\sigma_{\bar pp}=107.3\pm1.1$ mb and $\rho_{\bar pp}=0.132\pm 0.001$ for the Large Hadron Collider---robust predictions that rely critically  on the saturation of the Froissart bound.

\section*{Acknowledgments} The work of FH is supported  in part by the U.S.~Department of Energy under Grant No.~DE-FG02-95ER40896 and in part by the University of Wisconsin Research Committee with funds granted by the Wisconsin Alumni Research Foundation.

\newpage

\begin{table}[h,t]                   
%
\def\arraystretch{1.5}            
\begin{tabular}[b]{|l||c||c||}
    \cline{1-3}
      \multicolumn{1}{|c||}{Transition Energy Parameters}
      &\multicolumn{1}{c||}{$\pi^+p$ and $\pi^-p$ Scattering}
      &\multicolumn{1}{c||}{$pp$ and $\bar pp$  Scattering}\\
      \hline\hline
     $\nu_0$, lab transition energy \ \ \ \ \ \ \ \ \ \  (GeV)&3.12&7.59  					\\ 		 
	$\rightarrow \ \sqrt s_0$, cms transition energy \ \ (GeV)&2.6&4\\
		\hline
	$\sigma^+(\nu_0)$ \ \ \ \ \ \ \ \ \ \  \    (mb)&28.91&40.18\\
      $\sigma^-(\nu_0)$  \ \ \ \ \ \ \ \ \ \  \   (mb)&32.04&56.99\\

      $\left(\frac{d\sigma^+}{d(\nu/m)}\right)_{\nu=\nu_0}$\ \ \ \ (mb)
		&-0.2305 &-0.2262	\\ 
      $\left(\frac{d\sigma^-}{d(\nu/m)}\right)_{\nu=\nu_0}$\ \ \ \ (mb)
	    &-1.446&-0.2740\\
      \hline\hline\hline
\multicolumn{1}{|c||}{Minimum fitting energy} &\multicolumn{1}{c}{}&\multicolumn{1}{c||}{}\\
\hline
$\nu_{\rm min}$,\ \ \  \ \ \ \ \ lab minimum energy \ (GeV)&18.71&18.25\\
$\rightarrow \ \sqrt s_{\rm min}$, cms minimum energy\   (GeV)&6.0&6.0\\
     \hline
\multicolumn{3}{c}{$m$ is the pion (proton) mass and $\nu$ is the laboratory pion (proton) energy}
\end{tabular}
     \caption{\protect\small The transition energy parameters used for fitting  $\pi^+p$, $\pi^-p$ , $pp$ and  $\bar pp $ scattering.\label{table:transitionparameters}
}
\end{table}

\begin{table}[h,t]                   
%
\def\arraystretch{1.5}            
\begin{tabular}[b]{|l||c|c||c||}
\cline{2-4}
\multicolumn{1}{c|}{}&\multicolumn{2}{c||}{$\sigma\sim \ln^2(\nu/m_\pi)$}&\multicolumn{1}{c||}{$\sigma\sim \ln(\nu/m_\pi)$}\\
\cline{1-1}
\multicolumn{1}{|c||}{Parameters }
      &\multicolumn{2}{|c||}{$\delchimax$}&\multicolumn{1}{|c||}{$\delchimax$}\\ 
\cline{2-4}
	\multicolumn{1}{|c||}{}
      &\multicolumn{1}{c|}{6}&\multicolumn{1}{c||}{9} &\multicolumn{1}{c||}{6}\\
      \hline
	\multicolumn{4}{|c||}{\ \ \ \ \  Even Amplitude}\\
	\cline{1-4}
      $c_0$\ \ \   (mb)&$20.11$ &$20.32$&12.75\\ 
      $c_1$\ \ \   (mb)&$-0.921\pm0.110$ &$-0.981\pm0.100$&$1.286\pm 0.0056$\\ 
	$c_2$\ \ \ \   (mb)&$0.1767\pm0.0085$&$0.1815\pm0.0077$&------\\
      $\beta_{\cal P'}$\ \   (mb)&$54.40$ &$54.10$&64.87\\ 
      $\mu$&$0.5$ &$0.5$&0.5\\ 
	$f(0)$ (mb GeV)&$-2.33\pm0.36$&$-2.31\pm 0.35$&$0.34\pm 0.36$\\
      \hline
	\multicolumn{4}{|c||}{\ \ \ \ \  Odd Amplitude}\\
	\hline
      $\delta$\ \ \   (mb)&$-4.51$ &$-4.51$&-4.51\\
      $\alpha$&$0.660$ &$0.660$&0.660\\ 
	\cline{1-4}
     	\hline
	\hline
	$\chi^2_{\rm min}$&148.1&204.4&941.8\\
	${\cal R}\times\chi^2_{\rm min}$&164.3&210.0&1044.9\\ 
	$\nu$ (d.f).&127&135&128\\
\hline
	${\cal R}\times\chi^2_{\rm min}/\nu$&1.294&1.555&8.163\\
\hline
\end{tabular}
     \caption{\protect\small The fitted results for a 3-parameter $\chi^2$ fit with $\sigma\sim\ln^2(\nu/m_\pi)$ and a 2-parameter fit with $\sigma\sim\ln(\nu/m_\pi)$ to the total cross sections and $\rho$-values for $\pi^+ p$ and $\pi^- p$ scattering. The renormalized $\chi^2/\nu_{\rm min}$,  taking into account the effects of the $\delchimax$ cut, is given in the row  labeled ${\cal R}\times\chi^2_{\rm min}/\nu$. The errors in the fitted parameters have been multiplied by the appropriate $r_{\chi2}$. The pion mass  is $m_\pi$ and the laboratory pion energy is $\nu$. \label{table:pipfitnew}}
\end{table}%
\def\arraystretch{1}  
 
\begin{table}[h,t]                   
%
\def\arraystretch{1.5}            
\begin{tabular}[b]{|l||c|c||c||}
\cline{2-4}
\multicolumn{1}{c|}{}&\multicolumn{2}{c||}{$\sigma\sim \ln^2(\nu/m_p)$}&\multicolumn{1}{c||}{$\sigma\sim \ln(\nu/m_p)$}\\
\cline{1-1}
\multicolumn{1}{|c||}{Parameters }
      &\multicolumn{2}{|c||}{$\delchimax$}&\multicolumn{1}{|c||}{$\delchimax$}\\ 
\cline{2-4}
	\multicolumn{1}{|c||}{}
      &\multicolumn{1}{c|}{6}&\multicolumn{1}{c||}{9} &\multicolumn{1}{c||}{6}\\
      \hline
	\multicolumn{4}{|c||}{\ \ \ \ \  Even Amplitude}\\
	\cline{1-4}
      $c_0$\ \ \   (mb)&$37.32$ &$37.25$&28.26\\ 
      $c_1$\ \ \   (mb)&$-1.440\pm0.070$ &$-1.416\pm0.066$&$2.651\pm 0.0070$\\ 
	$c_2$\ \ \ \   (mb)&$0.2817\pm0.0064$&$0.2792\pm0.0059$&------\\
      $\beta_{\cal P'}$\ \   (mb)&$37.10$ &$37.17$&47.98\\ 
      $\mu$&$0.5$ &$0.5$&0.5\\ 
	$f(0)$ (mb GeV)&$-0.075\pm0.59$&$-0.069\pm 0.57$&$4.28\pm 0.59$\\
      \hline
	\multicolumn{4}{|c||}{\ \ \ \ \  Odd Amplitude}\\
	\hline
      $\delta$\ \ \   (mb)&$-28.56$ &$-28.56$&-28.56\\
      $\alpha$&$0.415$ &$0.415$&0.415\\ 
	\cline{1-4}
     	\hline
	\hline
	$\chi^2_{\rm min}$&181.6&216.6&2355.7\\
	${\cal R}\times\chi^2_{\rm min}$&201.5&222.5&2613.7\\ 
	$\nu$ (d.f).&184&189&185\\
\hline
	${\cal R}\times\chi^2_{\rm min}/\nu$&1.095&1.178&14.13\\
\hline
\end{tabular}
     \caption{\protect\small The fitted results for a 3-parameter $\chi^2$ fit with $\sigma\sim\ln^2(\nu/m_p)$ and a 2-parameter fit with $\sigma\sim\ln(\nu/m_p)$ to the total cross sections and $\rho$-values for $pp$ and $\bar pp$ scattering. The renormalized $\chi^2/\nu_{\rm min}$,  taking into account the effects of the $\delchimax$ cut, is given in the row  labeled ${\cal R}\times\chi^2_{\rm min}/\nu$. The errors in the fitted parameters have been multiplied by the appropriate $r_{\chi2}$.  The proton mass  is $m_p$ and the laboratory nucleon energy is $\nu$. \label{table:ppfitnew}}
\end{table}
\def\arraystretch{1}  
\begin{table}[h,t]                   
%
\def\arraystretch{1.5}            
\begin{tabular}[b]{|l||c|c||c|c||}
    \cline{1-5}
      \multicolumn{1}{|l||}{ $\sqrt s$, in GeV}
      &\multicolumn{1}{c|}{$\sigma_{\bar pp}$, in mb}
      &\multicolumn{1}{c||}{$\rho_{\bar p p}$}&\multicolumn{1}{c|}{$\sigma_{ pp}$, in mb}&\multicolumn{1}{c||}{$\rho_{pp}$}\\

      \hline\hline
	6&$48.97\pm0.01$&$-0.087\pm0.008$&$38.91\pm0.01$&$-.307\pm0.001$\\\hline60&$43.86\pm0.04$&$0.089\pm0.001$&$43.20\pm0.04$&$0.079\pm0.001$\\\hline
	100&$46.59\pm0.08$&$0.108\pm0.001$&$46.23\pm0.08$&$0.103\pm 0.001$\\\hline
	300&$55.03\pm0.21$&$0.131\pm0.001$&$54.93\pm0.21$&$0.130\pm 0.002$\\\hline	
	400&$57.76\pm0.25$&$0.134\pm0.002$&$57.68\pm0.25$&$0.133\pm 0.002$\\\hline
	540&$60.81\pm0.29$&$0.137\pm0.002$&$60.76\pm0.29$&$0.136\pm0.002$\\\hline
 	1,800&$75.19\pm0.55$&$0.139\pm0.001$&$75.18\pm0.55$&$0.139\pm0.001$\\\hline    
 	14,000&$107.3\pm1.2$&$0.132\pm0.001$&$107.3\pm1.2$&$0.132\pm0.001$\\\hline    
 	16,000&$109.8\pm1.3$&$0.131\pm0.001$&$109.8\pm1.3$&$0.131\pm0.001$\\\hline   	
 	50,000&$132.1\pm1.7$&$0.124\pm0.001$&$132.1\pm1.7$&$0.124\pm0.001$\\\hline
 	100,000&$147.1\pm2.0$&$0.120\pm0.001$&$147.1\pm2.0$&$0.120\pm0.001$\\\hline
\end{tabular}
     \caption{\protect\small Predictions of high energy $\bar pp$ and $pp$ total  cross sections and $\rho$-values,  from Table \ref{table:ppfitnew}, $\sigma\sim\ln^2(\nu/m_\pi)$, $\delchimax=6$.\label{table:predictions}
}
\end{table}

\def\arraystretch{1}  
\def\arraystretch{1}  

\begin{figure}[h,t,b] 
\begin{center}
\mbox{\epsfig{file=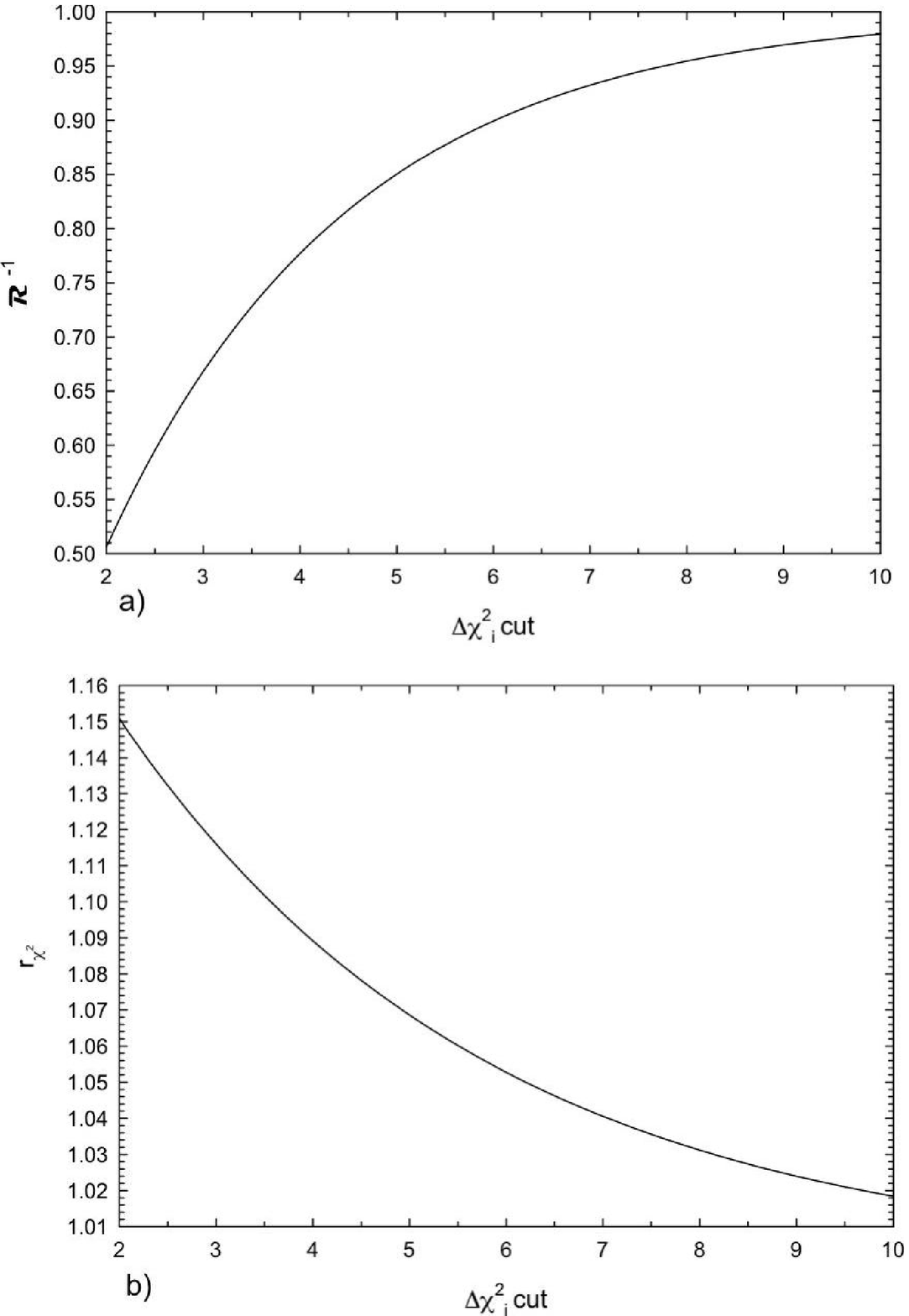,width=4.8in%
,bbllx=65pt,bblly=135pt,bburx=510pt,bbury=775pt,clip=%
}}
\end{center}
\caption[]{ \footnotesize 
a) A plot of  ${\cal R}^{-1}$, the reciprocal of the factor  that multiplies $\chi^2_{\rm min}/\nu$ found in the $\chi^2$ fit to the sifted data set  {\em vs.} $\delchi$ cut, {\em i.e.,} $\delchimax$.
\mbox{\ \ } b) A plot of $r_{\chi^2}$, the  factor whose square multiplies the covariant matrix found in the $\chi^2$ fit to the sifted data set  {\em vs.} $\delchi$ cut, {\em i.e.,} $\delchimax$. These figures are taken from ref. \cite{sieve}.  
  }
\label{renorm}
\end{figure}

\begin{figure}[h,t,b] 
\begin{center}
\mbox{\epsfig{file=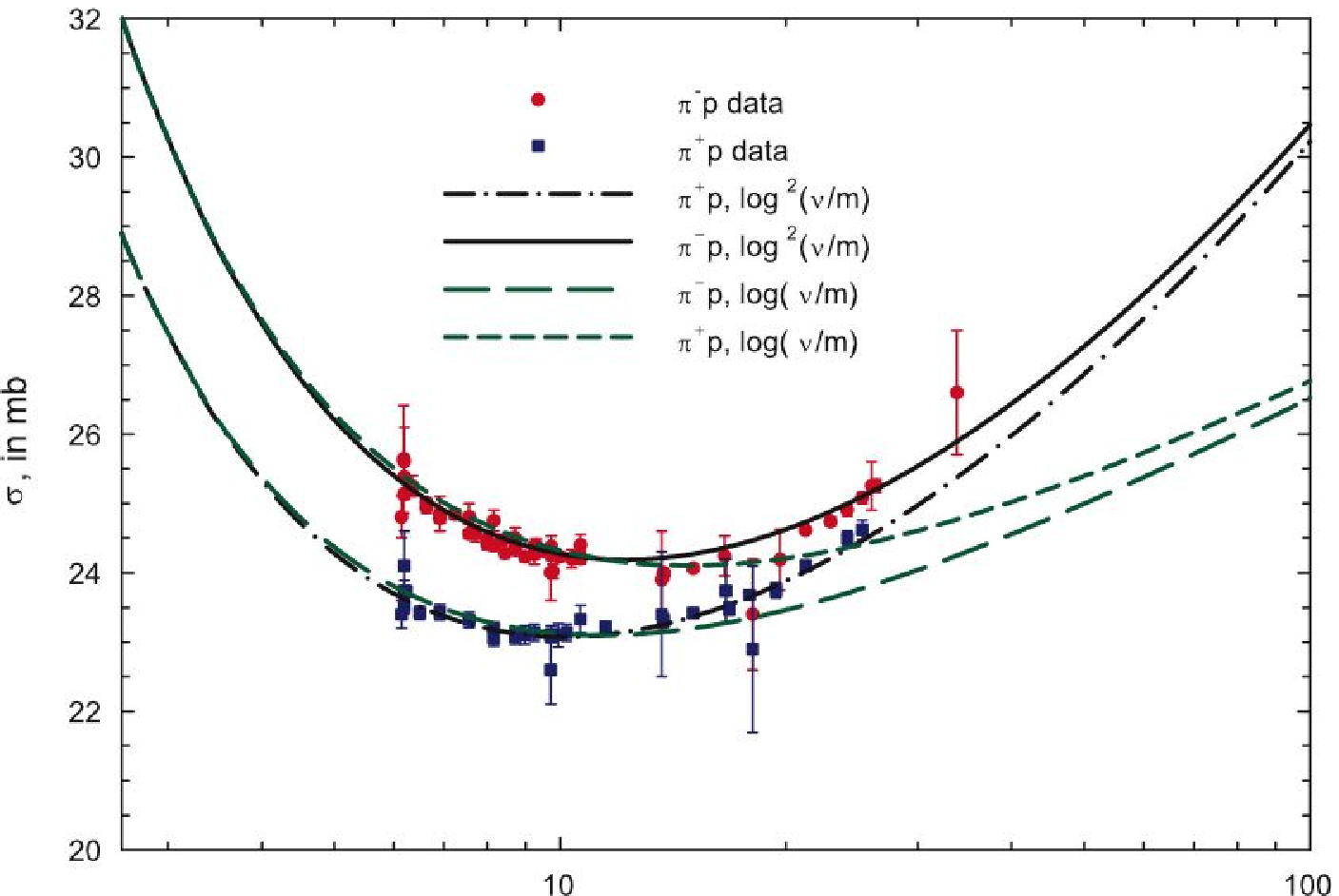
,width=6in,
bbllx=0pt,bblly=0pt,bburx=420pt,bbury=320pt,clip=%
}}
\end{center}
\caption[]{ \footnotesize
The fitted total cross sections $\sigma_{\pi^+p}$ and $\sigma_{\pi^-p}$ in mb, {\em vs.} $\sqrt s$, in GeV, using the 4 constraints of Equations (\ref{deriveven}), (\ref{intercepteven}), (\ref{derivodd}) and (\ref{interceptodd}).  The circles are the sieved  data  for $\pi^-p$ scattering and the squares are the sieved data for $\pi^+p$ scattering for $\sqrt s\ge 6$ GeV. The dash-dotted curve ($\pi^+ p$)  and the solid curve ($\pi^- p$) are $\chi^2$ fits (Table \ref{table:pipfitnew}, $\sigma\sim\ln^2(\nu/m_\pi)$, $\delchimax=6$) of  the high energy data  of the form~: $\sigma_{\pi^{\pm}p}=c_0 +c_1{\ln }\left({\nu\over m}\right)+c_2{\ln }^2\left({\nu\over m}\right)+\beta_{\cal P'}\left({\nu\over m}\right)^{\mu -1}\pm \delta\left({\nu\over m}\right)^{\alpha -1}$. The upper sign is for $\pi^+p$ and the lower sign is for $\pi^-p$ scattering.  The short dashed curve ($\pi^+ p$) and the long dashed curve ($\pi^- p$) are $\chi^2$ fits (Table  \ref{table:pipfitnew}, $\sigma\sim\ln(\nu/m_\pi)$, $\delchimax=6$ ) of  the high energy data  of the form~: $\sigma_{\pi^{\pm}p}=c_0 +c_1{\ln }\left({\nu\over m}\right)+\beta_{\cal P'}\left({\nu\over m}\right)^{\mu -1}\pm\delta\left({\nu\over m}\right)^{\alpha -1}$. The upper sign is for $\pi^+p$ and the lower sign is for $\pi^-p$ scattering. The laboratory energy of the pion is  $\nu$ and $m$ is the pion mass. 
}
\label{fig:sigpip}
\end{figure}
\begin{figure}[h,t,b] 
\begin{center}
\mbox{\epsfig{file=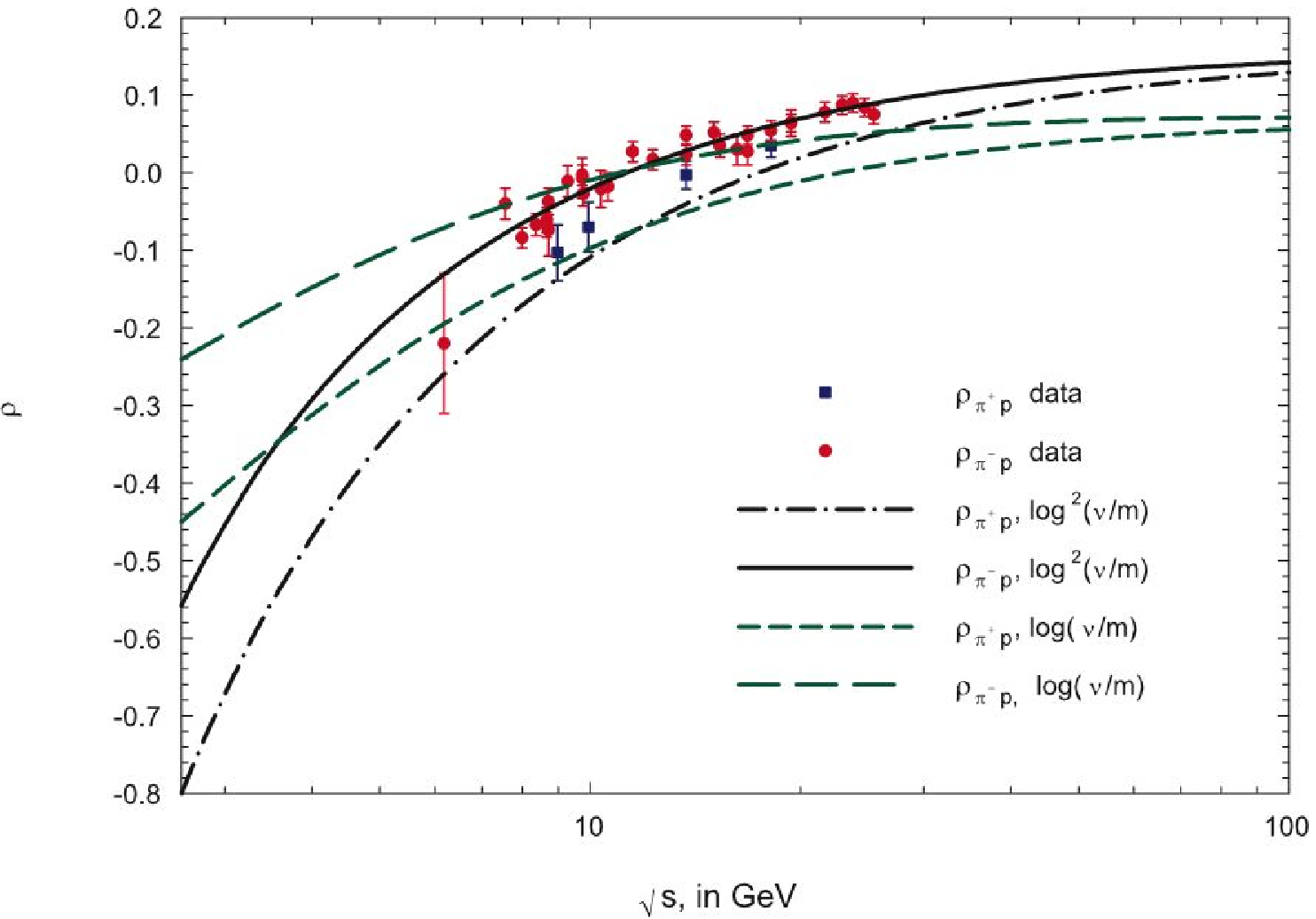,width=6in%
,bbllx=0pt,bblly=0pt,bburx=460pt,bbury=314pt,clip=%
}}
\end{center}
\caption[]{ \footnotesize
The fitted $\rho$-values, $\rho_{\pi^+p}$ and $\rho_{\pi^-p}$, {\em vs.} $\sqrt s$, in GeV, using the 4 constraints of Equations (\ref{deriveven}), (\ref{intercepteven}), (\ref{derivodd}) and (\ref{interceptodd}).  The circles are the sieved data  for $\pi^-p$ scattering and the squares are the sieved data for $\pi^+p$ scattering for $\sqrt s\ge 6$ GeV. The  dash-dotted curve ($\pi^+ p$) and the solid curve ($\pi^- p$) are $\chi^2$ fits (Table \ref{table:pipfitnew}, $\sigma\sim\ln^2(\nu/m_\pi)$, $\delchimax=6$) of  the high energy data  of the form~: $\rho^\pm= {1\over\sigma^\pm} \left\{\frac{\pi}{2}c_1+ c_2\pi\ln\left(\frac{\nu}{m}\right)-\beta_{\cal P'}\cot(\pi\mu/2)\left(\frac{\nu}{m}\right)^{\mu -1}+\frac{4\pi}{\nu}f_+(0)\ \pm \ \delta\tan(\pi\alpha/ 2)\left({\nu\over m}\right)^{\alpha -1}\right\} $. The upper sign is for $\pi^+p$ and the lower sign is for $\pi^-p$ scattering.  The short dashed curve ($\pi^+ p$) and the long dashed curve ($\pi^- p$)  are $\chi^2$ fits (Table  \ref{table:pipfitnew}, $\sigma\sim\ln(\nu/m_\pi)$, $\delchimax=6$ ) of  the high energy data  of the form~: $\rho^\pm= {1\over\sigma^\pm} \left\{\frac{\pi}{2}c_1-\beta_{\cal P'}\cot(\pi\mu/2)\left(\frac{\nu}{m}\right)^{\mu -1}+\frac{4\pi}{\nu}f_+(0)\ \pm \ \delta\tan(\pi\alpha/ 2)\left({\nu\over m}\right)^{\alpha -1}\right\}$. The upper sign is for $\pi^+p$ and the lower sign is for $\pi^-p$ scattering.  The laboratory energy of the pion is  $\nu$ and $m$  is the pion mass.
  }
\label{fig:rhopip}
\end{figure}
%
\begin{figure}[h,t,b] 
\begin{center}
\mbox{\epsfig{file=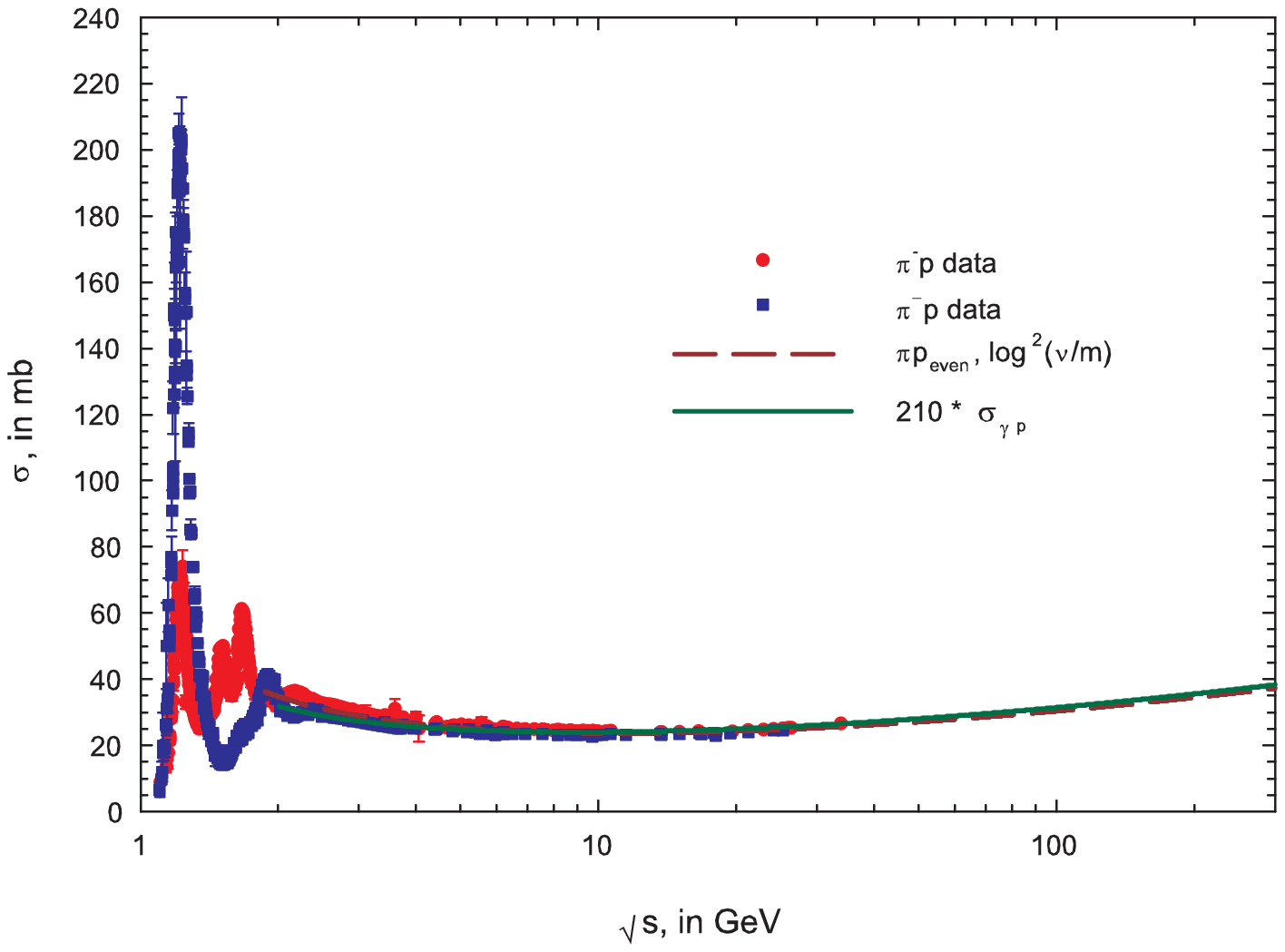,width=6in%
,bbllx=0pt,bblly=0pt,bburx=430pt,bbury=340pt,clip=%
}}
\end{center}
\caption[]{ \footnotesize
The circles are the cross section data  for $\pi^-p$ scattering and the squares are the cross section data for $\pi^+p$ scattering, in mb, {\em vs.} $\sqrt s$, in GeV, for all of the known data.  The dashed curve is the $\chi^2$ fit (Table \ref{table:pipfitnew}, $\sigma\sim\ln^2(\nu/m_\pi)$, $\delchimax=6$) to  the high energy cross section data  of the even amplitude cross section, of the form~: $\sigma_{\pi p^{\rm even}}=c_0 +c_1{\ln }\left({\nu\over m}\right)+c_2{\ln }^2\left({\nu\over m}\right)+\beta_{\cal P'}\left({\nu\over m}\right)^{\mu -1}$, with  $c_0$ and $\beta_{\cal P'}$ constrained by \eq{deriveven} and \eq{intercepteven}.  
The laboratory energy of the pion is  $\nu$ and $m$ is the pion mass. The dashed curve is  $210\times \sigma_{\gamma p}$, from a fit of $\gamma p$ cross sections by Block and Halzen\cite{BH}  of the form: $\sigma_{\gamma p}=c_0 +c_1{\ln }(\nu/m_p)+c_2{\ln }^2(\nu/m_p)+\beta_{\cal P'}/\sqrt{\nu/m_p}$, where $m_p$ is the proton mass.  The $\gamma p$ cross sections were fit for  cms energies $\sqrt s\ge2.01$ GeV, whereas the $\pi p$ data (cross sections and $\rho$-values) were fit for cms energies $\sqrt s\ge6$ GeV. The two fitted curves are virtually indistinguishable in the energy region $2\le \sqrt s\le 300$ GeV.
}
\label{fig:sigpipallenergies}
\end{figure}
%
\begin{figure}[h,t,b] 
\begin{center}
\mbox{\epsfig{file=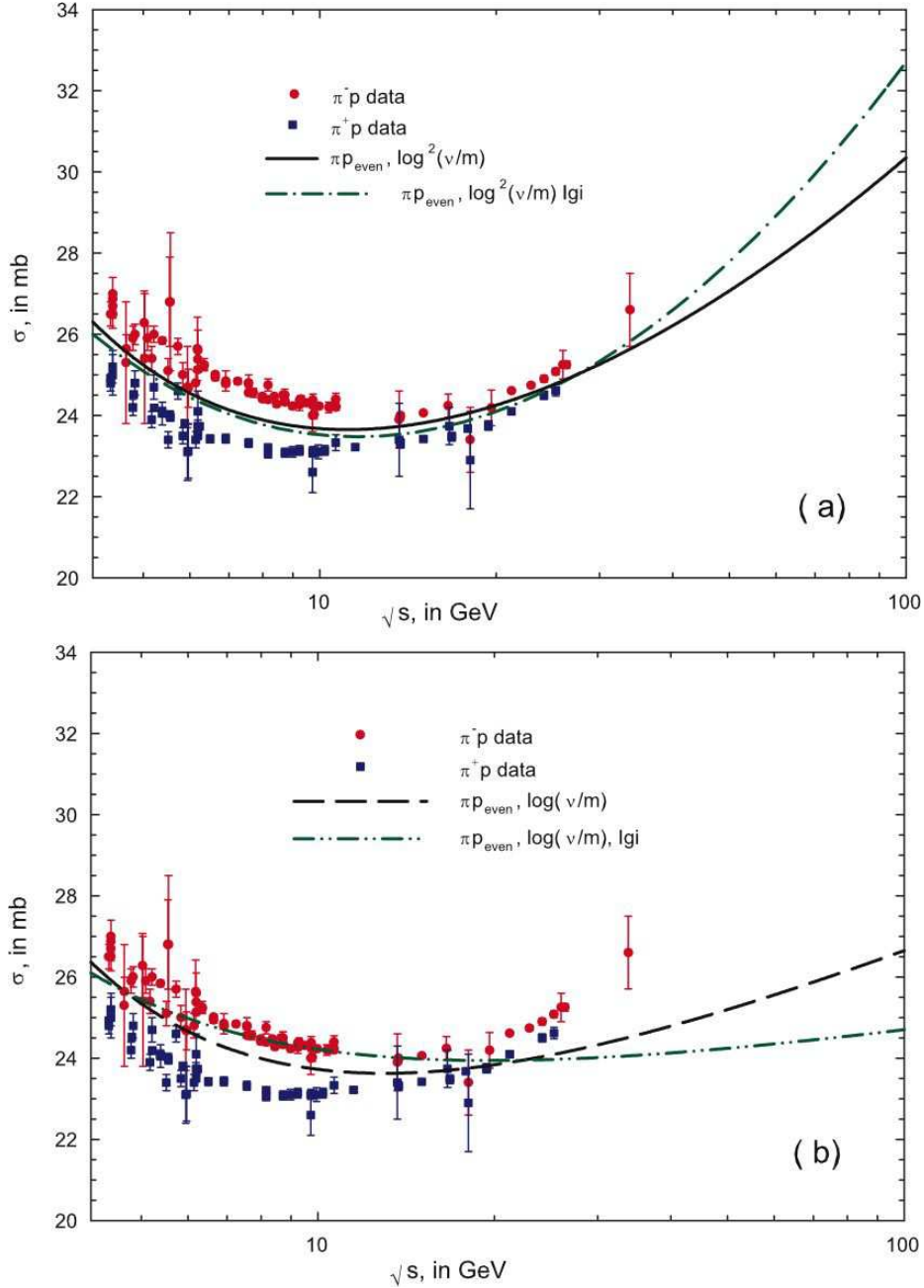,width=5in%
,bbllx=0pt,bblly=0pt,bburx=420pt,bbury=595pt,clip=%
}}
\end{center}
\caption[]{ \footnotesize
The circles are the cross section data  for $\pi^-p$ scattering and the squares are the cross section data for $\pi^+p$ scattering for all known data, {\em vs.} $\sqrt s$, in GeV. The solid curve in Fig.(a) is the $\chi^2$ fit (Table \ref{table:pipfitnew}, $\sigma\sim\ln^2(\nu/m_\pi)$, $\delchimax=6$) to  the high energy cross section data  of the even amplitude, of the form~: $\sigma_{\pi p^{\rm even}}=c_0 +c_1{\ln }\left({\nu\over m}\right)+c_2{\ln }^2\left({\nu\over m}\right)+\beta_{\cal P'}\left({\nu\over m}\right)^{\mu -1}$, with  $c_0$ and $\beta_{\cal P'}$ constrained by \eq{deriveven} and \eq{intercepteven}.  The dash-dotted curve is an even amplitude $\ln^2(\nu/m_\pi)$ fit made by Igi and Ishida\cite{igi}, using finite energy sum rules (FESR). The dashed curve in Fig.(b) is the $\chi^2$ fit (  Table \ref{table:pipfitnew}, $\sigma\sim\ln(\nu/m_\pi)$, $\delchimax=6$) to  the high energy cross section data  of the even amplitude, of the form~: $\sigma_{\pi p^{\rm even}}=c_0 +c_1{\ln }\left({\nu\over m}\right))+\beta_{\cal P'}\left({\nu\over m}\right)^{\mu -1}$, with  $c_0$ and $\beta_{\cal P'}$ constrained by \eq{deriveven} and \eq{intercepteven}.  The dot-dot-dashed curve is an even amplitude $\ln(\nu/m_\pi)$ fit made by Igi and Ishida\cite{igi}, using finite energy sum rules. 
The laboratory energy of the pion is  $\nu$ and $m$ is the pion mass. 
}
\label{fig:blockhalzen&Igi}
\end{figure}
\begin{figure}[h,t,b] 
\begin{center}
\mbox{\epsfig{file=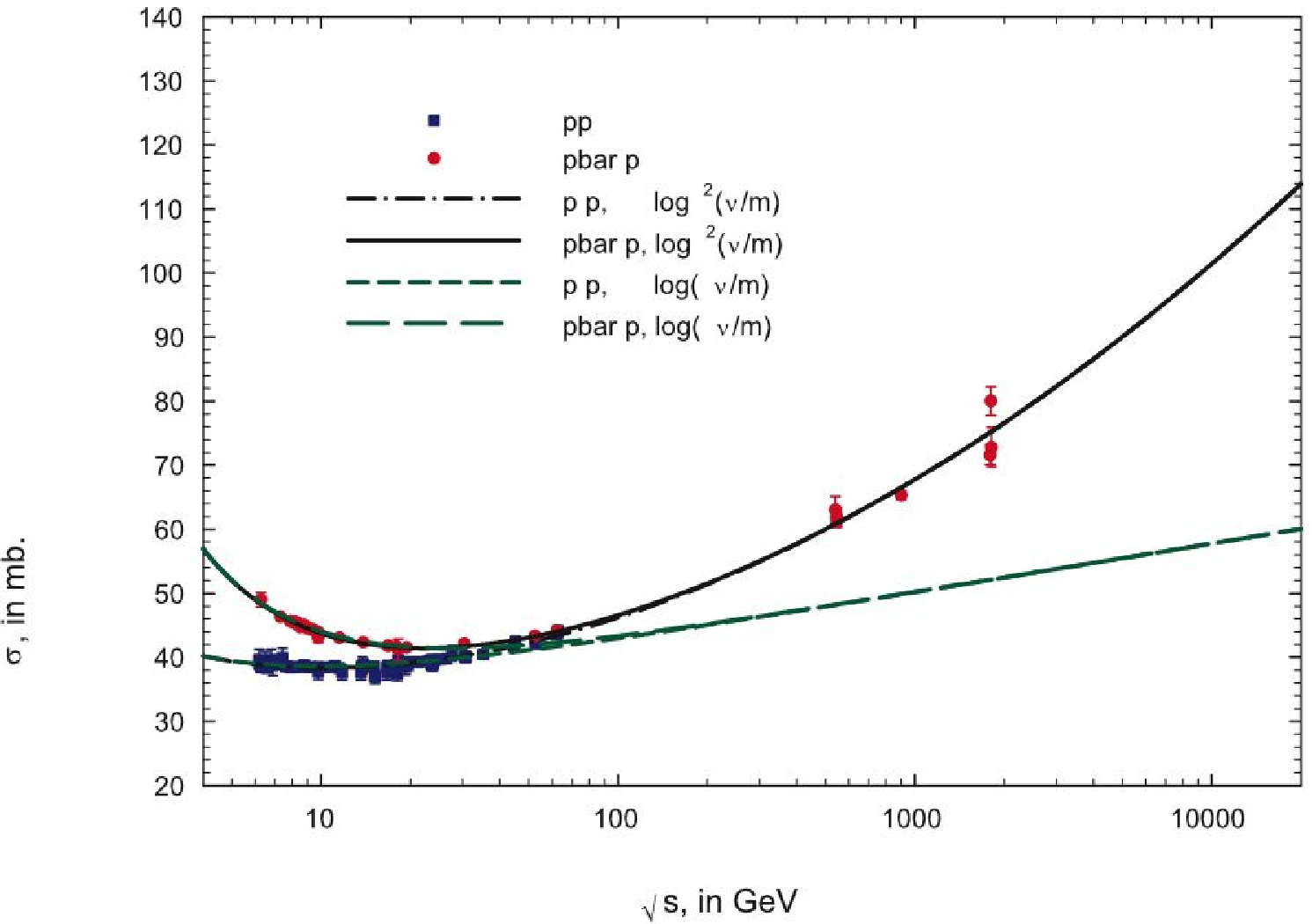,width=6in%
,bbllx=0pt,bblly=0pt,bburx=445pt,bbury=535pt,clip=%
}}
\end{center}
\caption[]{ \footnotesize
The fitted total cross sections $\sigma_{p p}$ and $\sigma_{\pbar p}$ in mb, {\em vs.} $\sqrt s$, in GeV, using the 4 constraints of Equations (\ref{deriveven}), (\ref{intercepteven}), (\ref{derivodd}) and (\ref{interceptodd}).  The circles are the sieved data  for $\pbar p$ scattering and the squares are the sieved data for $p p$ scattering for $\sqrt s\ge 6$ GeV. The dash-dotted curve ($pp$)  and the solid curve ($\pbar p$) are $\chi^2$ fits (Table \ref{table:ppfitnew}, $\sigma\sim\ln^2(\nu/m_\pi)$, $\delchimax=6$) of  the high energy data  of the form~: $\sigma^\pm=c_0 +c_1{\ln }\left({\nu\over m}\right)+c_2{\ln }^2\left({\nu\over m}\right)+\beta_{\cal P'}\left({\nu\over m}\right)^{\mu -1}\pm \delta\left({\nu\over m}\right)^{\alpha -1}$. The upper sign is for $p p$ and the lower sign is for $\pbar p$ scattering.  The short dashed curve ($p p$) and the long dashed curve ($\pbar p$) are $\chi^2$ fits (Table  \ref{table:ppfitnew}, $\sigma\sim\ln(\nu/m_\pi)$, $\delchimax=6$ ) of  the high energy data  of the form~: $\sigma^\pm=c_0 +c_1{\ln }\left({\nu\over m}\right)+\beta_{\cal P'}\left({\nu\over m}\right)^{\mu -1}\pm\delta\left({\nu\over m}\right)^{\alpha -1}$. The upper sign is for $p p$ and the lower sign is for $\pbar p$ scattering. The laboratory energy of the nucleon is  $\nu$ and $m$ is the nucleon mass. 
}
\label{fig:sigmapp}
\end{figure}
%
\begin{figure}[h,t,b] 
\begin{center}
\mbox{\epsfig{file=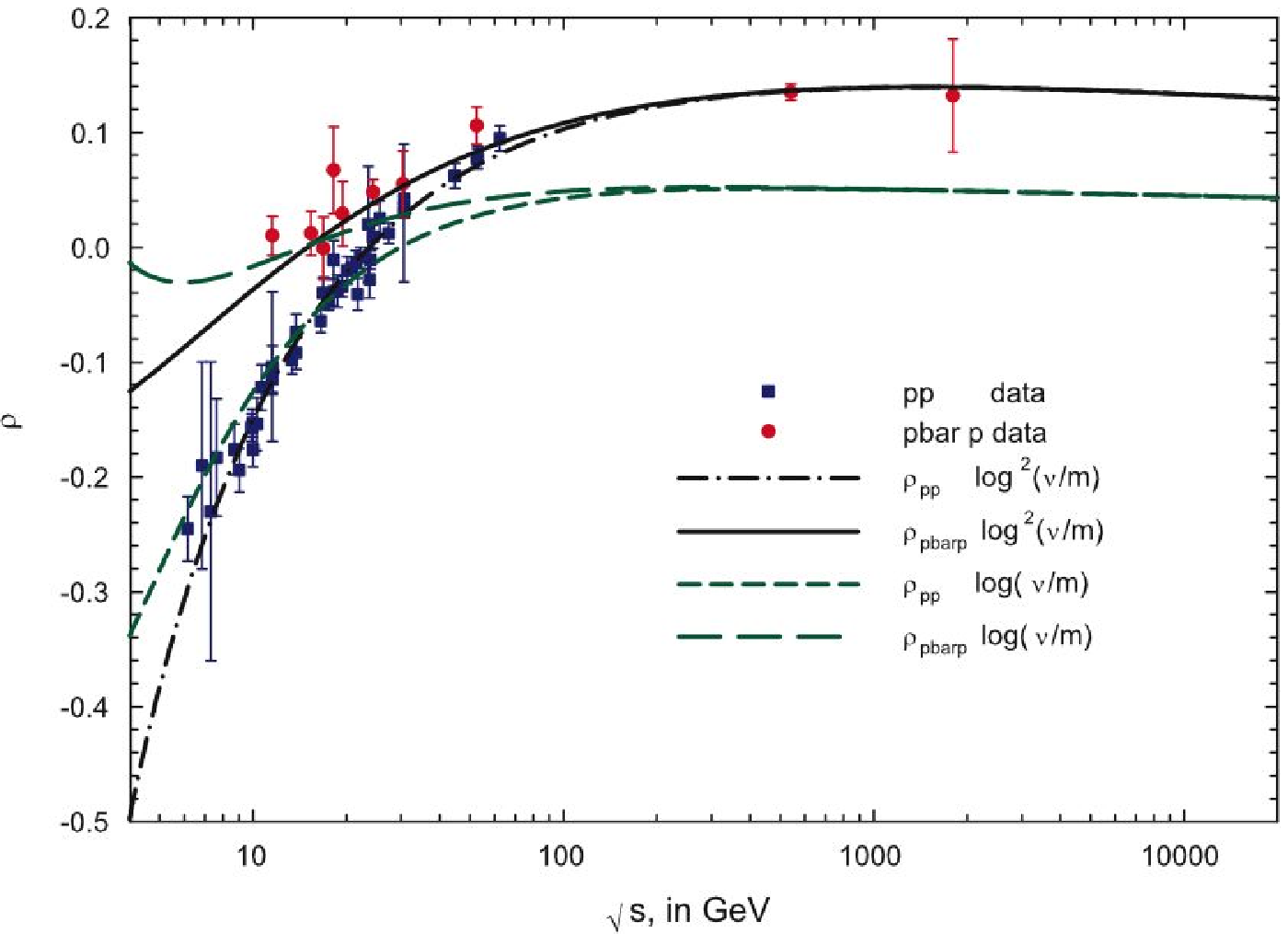,width=6in%
,bbllx=0pt,bblly=0pt,bburx=435pt,bbury=305pt,clip=%
}}
\end{center}
\caption[]{ \footnotesize
The fitted $\rho$-values, $\rho_{p p}$ and $\rho_{\pbar p}$, {\em vs.} $\sqrt s$, in GeV, using the 4 constraints of Equations (\ref{deriveven}), (\ref{intercepteven}), (\ref{derivodd}) and (\ref{interceptodd}).  The circles are the sieved data  for $\pbar p$ scattering and the squares are the sieved data for $p p$ scattering for $\sqrt s\ge 6$ GeV. The  dash-dotted curve ($p p$) and the solid curve ($\pbar p$) are $\chi^2$ fits (Table \ref{table:ppfitnew}, $\sigma\sim\ln^2(\nu/m_\pi)$, $\delchimax=6$) of  the high energy data  of the form~: $\rho^\pm= {1\over\sigma^\pm} \left\{\frac{\pi}{2}c_1+ c_2\pi\ln\left(\frac{\nu}{m}\right)-\beta_{\cal P'}\cot(\pi\mu/2)\left(\frac{\nu}{m}\right)^{\mu -1}+\frac{4\pi}{\nu}f_+(0)\ \pm \ \delta\tan(\pi\alpha/ 2)\left({\nu\over m}\right)^{\alpha -1}\right\} $. The upper sign is for $p p$ and the lower sign is for $\pbar p$ scattering.  The short dashed curve ($p p$) and the long dashed curve ($\pbar p$)  are $\chi^2$ fits (Table  \ref{table:ppfitnew}, $\sigma\sim\ln(\nu/m_\pi)$, $\delchimax=6$ ) of  the high energy data  of the form~: $\rho^\pm= {1\over\sigma^\pm} \left\{\frac{\pi}{2}c_1-\beta_{\cal P'}\cot(\pi\mu/2)\left(\frac{\nu}{m}\right)^{\mu -1}+\frac{4\pi}{\nu}f_+(0)\ \pm \ \delta\tan(\pi\alpha/ 2)\left({\nu\over m}\right)^{\alpha -1}\right\}$. The upper sign is for $p p$ and the lower sign is for $\pbar p$ scattering.  The laboratory energy of the nucleon is  $\nu$ and $m$  is the nucleon mass.
  }
\label{fig:rhopp}
\end{figure}
%
\begin{figure}[h,t,b] 
\begin{center}
\mbox{\epsfig{file=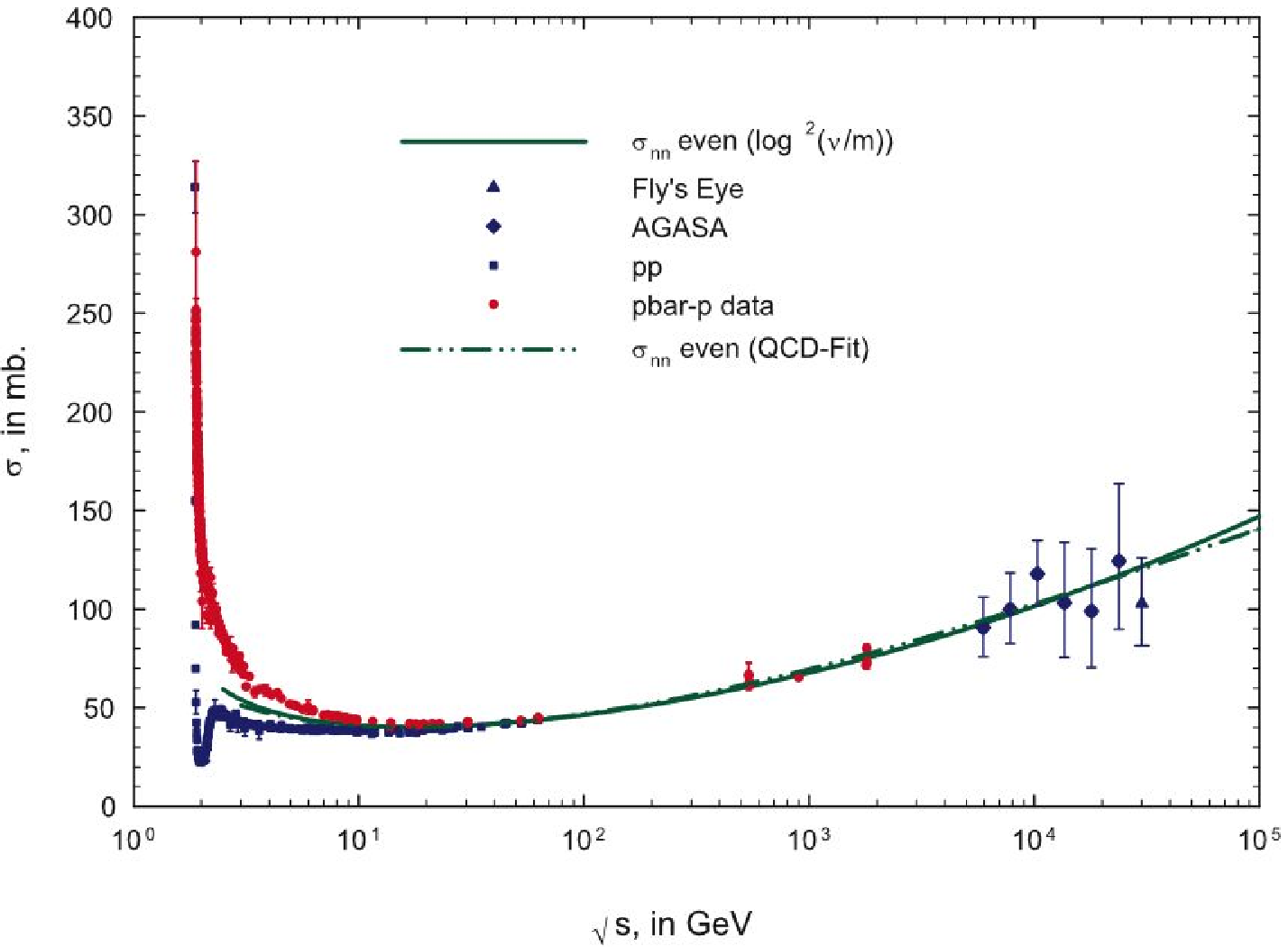,width=6in%
,bbllx=0pt,bblly=0pt,bburx=425pt,bbury=335pt,clip=%
}}
\end{center}
\caption[]{ \footnotesize
The circles are the cross section data  for $\pbar p$ scattering and the squares are the cross section data for $p p$ scattering, in mb, {\em vs.} $\sqrt s$, in GeV, for all of the known accelerator data. The solid  curve is the $\chi^2$ fit ((Table \ref{table:ppfitnew}, $\sigma\sim\ln^2(\nu/m_\pi)$, $\delchimax=6$) of  the high energy data  of the crossing-even amplitude, of the form~: $\sigma_{nn}{\rm even}=c_0 +c_1{\ln }\left({\nu\over m}\right)+c_2{\ln }^2\left({\nu\over m_p}\right)+\beta_{\cal P'}\left({\nu\over m}\right)^{\mu -1}$, with  $c_0$ and $\beta_{\cal P'}$ constrained by \eq{deriveven} and \eq{intercepteven}.    The dot-dot-dashed curve is the crossing-even amplitude cross section $\sigma_{nn}$, from a QCD-inspired fit that fit not only  the accelerator $\bar pp$ and $pp$ cross sections and $\rho$-values, but also fit the AGASA and Fly's Eye cosmic ray pp cross sections shown in the figure---work done several years ago by the Block, Halzen and Stanov (BHS group)\cite{BHS}. 
The laboratory energy of the proton is  $\nu$ and $m$ is the proton mass. It is most striking that the two fitted curves for $\sigma_{nn}$even, using on the one hand, the $\ln^2(\nu/m)$ model of this work and on the other hand, the QCD-inspired model of the BHS group\cite{BHS}, are virtually indistinguishable over 5 decades of cms energy, {\em i.e.,} in the energy region $3\le \sqrt s\le 10^5$ GeV.
}
\label{fig:sigppallenergies}
\end{figure}

\end{document}